\documentclass[aps,pre,superscriptaddress,showpacs,floatfix,amsmath,amssymb,longbibliography,11pt]{revtex4-1}
\usepackage{amssymb}
\usepackage{bm}
\usepackage{url}
\usepackage{graphicx}

\usepackage{geometry}
\usepackage{tgtermes}

\begin{document}

\title{Time irreversibility and multifractality of power along single particle trajectories in turbulence\footnote{Version accepted for publication (postprint) on Phys. Rev. Fluids 2, 104604 -- Published 27 October 2017}}

\author{Massimo Cencini} \thanks{Corresponding author}
\email{massimo.cencini@cnr.it} \affiliation{Istituto dei Sistemi
  Complessi, CNR, Via dei Taurini 19, 00185 Rome, Italy and INFN ``Tor Vergata''}

\author{Luca Biferale}
\affiliation{Dipartimento di Fisica and INFN, Universit\`a di Roma ``Tor Vergata'', Via Ricerca Scientifica 1, 00133 Roma, Italy}

\author{Guido Boffetta}
\affiliation{Dipartimento di Fisica and INFN, Universit\`a di Torino, 
Via P. Giuria 1, 10125 Torino, Italy}

\author{Massimo De Pietro}
\affiliation{Dipartimento di Fisica and INFN, Universit\`a di Roma ``Tor Vergata'', Via Ricerca Scientifica 1, 00133 Roma, Italy}

\begin{abstract}
  The irreversible turbulent energy cascade epitomizes strongly
  non-equilibrium systems.  At the level of single fluid particles,
  time irreversibility is revealed by the asymmetry of the rate of
  kinetic energy change, the Lagrangian power, whose moments
  display a power-law dependence on the Reynolds number, as
  recently shown by Xu \textit{et al}. [H Xu \textit{et al}, Proc. Natl. Acad. Sci. U.S.A.
    \textbf{111}, 7558 (2014)].  Here Lagrangian power statistics are rationalized within the multifractal
  model of turbulence, whose predictions are shown to agree with
  numerical and empirical data.  Multifractal predictions are also
  tested, for very large Reynolds numbers, in dynamical models of the
  turbulent cascade, obtaining remarkably good agreement for
  statistical quantities insensitive to the asymmetry and,
  remarkably, deviations for those probing the asymmetry. These
  findings raise fundamental questions concerning time irreversibility
  in the infinite-Reynolds-number limit of the Navier-Stokes
  equations.
\end{abstract}


\maketitle

\section{Introduction}
\label{sec:1}
In nature, the majority of the processes
involving energy flow occur in nonequilibrium conditions from the
molecular scale of biology \cite{ritort2008nonequilibrium} to
astrophysics \cite{priest2012solar}. Understanding such nonequilibrium
processes is of great interest at both fundamental and applied levels,
from small-scale technology \cite{blickle2012realization} to climate
dynamics \cite{kleidon2010life}. A key aspect of nonequilibrium
systems is the behavior of fluctuations that markedly differ from
equilibrium ones.  As for the latter, detailed balance establishes
equiprobability of forward and backward transitions between any two
states, a statistical manifestation of time reversibility
\cite{onsager1931}, while, irreversibility of nonequilibrium
processes breaks detailed balance.  In three-dimensional (3D)
turbulence, a prototype of very far-from-equilibrium systems,
detailed balance breaks in a fundamental way \cite{rose1978}: It is
more probable to transfer energy from large to small scales than its
reverse.  Indeed, in statistically stationary turbulence, energy,
supplied at scale $L$ at rate $\epsilon$ ($\approx U_L^3/L$, $U_L$
being the root mean square single-point velocity), is transferred with
a constant flux approximately equal to $\epsilon$ up to the scale $\eta$, where it is
dissipated at the same rate $\epsilon$, even for vanishing viscosity
($\nu \to 0$) \cite{Frisch1995}. As a result, time reversibility,
formally broken by the viscous term, is not restored for $\nu\to 0$
\cite{falkovich2006}.  Time irreversibility is unveiled by the
asymmetry of two-point statistical observables. In particular, the
constancy of the energy flux directly implies, in the Eulerian frame,
a non vanishing third moment of longitudinal velocity difference
between two points at distance $r$ (the $\frac{4}{5}$ law
\cite{Frisch1995}) and, in the Lagrangian frame, a faster separation
of particle pairs backward than forward in time
\cite{falkovich2013single,jucha2014time}.

\begin{figure}[t!]
\centering
\includegraphics[width=0.35\columnwidth]{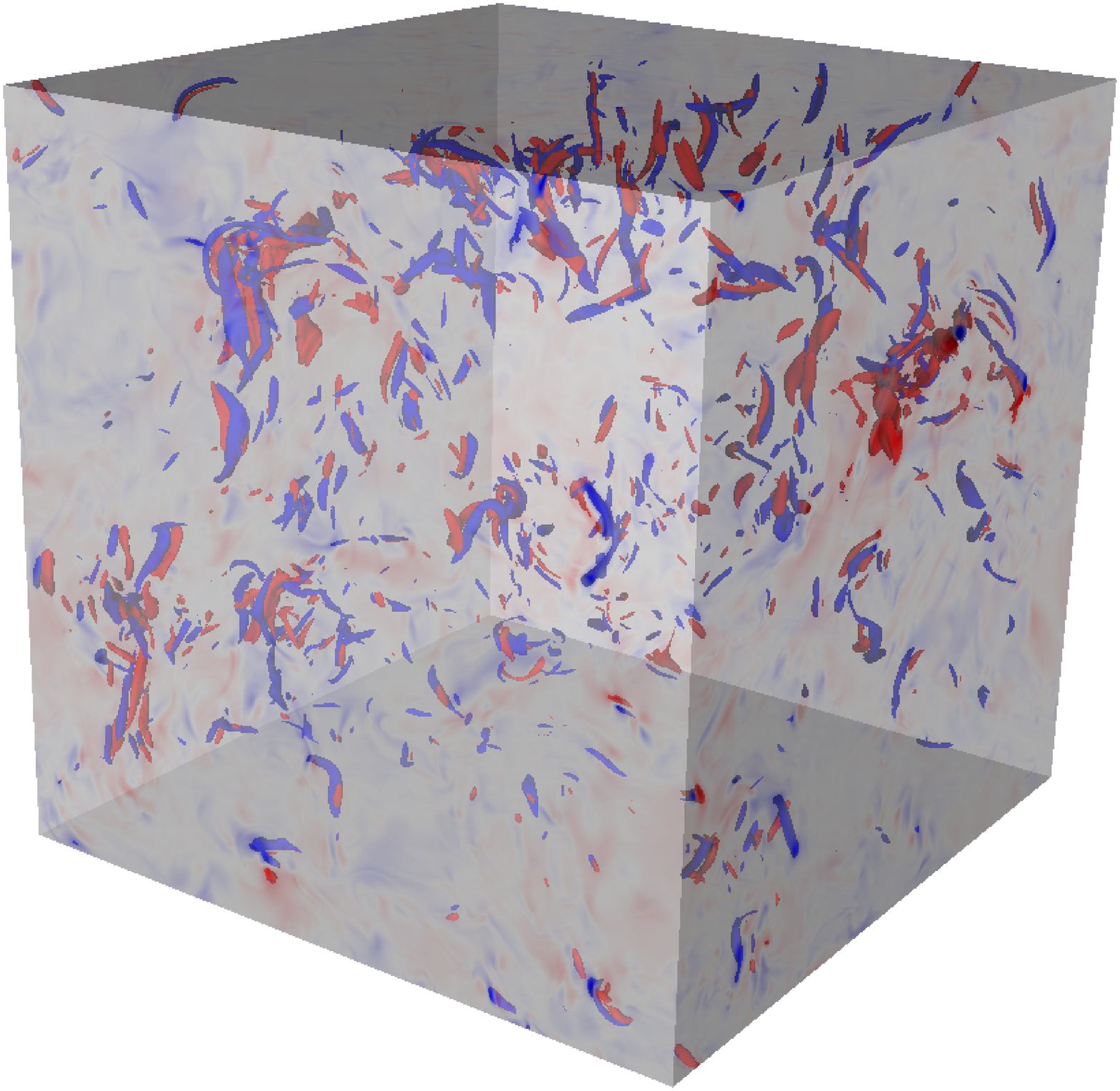}\hspace{0.5cm}
\includegraphics[width=0.42\columnwidth]{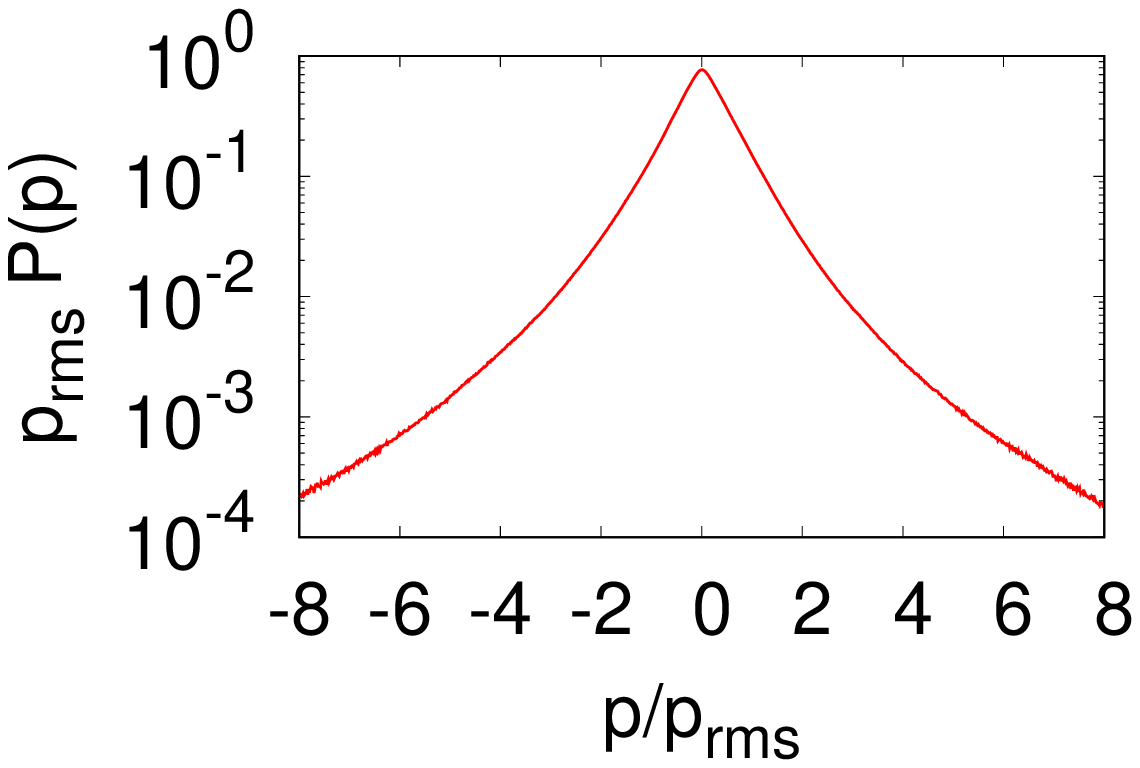}
\caption{Shown on the left is a three-dimensional rendering of the Lagrangian
  power spatial distribution in the whole simulation volume. Red (blue)
  represents the isosurfaces $p=\pm 6 p_{rms}$ ($p_{rms}=
  \langle p^2\rangle^{1/2}$), which appear clusterized in dipole
  structures. Shown on the right is the  log-lin standardized PDF of $p$ for $\mathrm{Re}_\lambda
  \approx 104$.  Notice that the asymmetry of the distribution is very
  small, hence the difficulty to quantify and rationalize the physics
  behind irreversible effects along a particle trajectory.
  \label{fig:p3d}}
\end{figure}

Remarkably, time irreversibility has been recently discovered
at the level of single-particle statistics \cite{xu2014,xu2014b} that
is not \textit{a priori} sensitive to the existence of a nonzero energy
flux. This opens important challenges also at applied levels for
stochastic modelization of single-particle transport, e.g., in
turbulent environmental flows \cite{sawford}.  Both experimental
and numerical data revealed that the temporal dynamics of Lagrangian
kinetic energy $E(t)=\frac{1}{2}v^2(t)$, where $\bm v(t)=\bm u(\bm
x(t),t)$ is the Lagrangian velocity along a particle trajectory $\bm
x(t)$, is characterized by events where $E(t)$ grows slower than it 
decreases.  Such {\it flight-crash} events result in the asymmetry of
distribution of the Lagrangian power, $p(t)=\dot{E}=\bm v(t)\cdot\bm
a(t)$ ($\bm a\equiv \dot{\bm v}=\partial_t \bm u+\bm u\cdot \bm \nabla
\bm u$ being the fluid particle acceleration). While in stationary
conditions the mean power vanishes $\langle p\rangle=0$, the third
moment is increasingly negative with the Taylor scale
Reynolds number $\mathrm{Re}_\lambda \approx (U_LL/\nu)^{1/2}\approx
T_L/\tau_\eta$ measuring the ratio between the timescales of energy
injection $T_L$ and dissipation $\tau_\eta$, which easily exceeds
$10^3$ in the laboratory.  In particular, it was found that $\langle
p^3\rangle/\epsilon^3 \sim -\mathrm{Re}_\lambda^{2}$ \cite{xu2014,xu2014b} and
$\langle p^2\rangle/\epsilon^2 \sim \mathrm{Re}_\lambda^{4/3}$.
Interestingly, the $\mathrm{Re}_\lambda$ dependence deviates from the
dimensional prediction based on Kolmogorov phenomenology \cite{Frisch1995}
$\langle p^q\rangle/\epsilon^q \propto
\mathrm{Re}_\lambda^{q/2}$, signaling that the Lagrangian power is strongly
intermittent as exemplified by its spatial distribution and the strong
non-Gaussian tails of the probability distribution function of $p$
(Fig.~\ref{fig:p3d}).

From a theoretical point of view, the above scaling behavior of the
power with $\mathrm{Re}_\lambda$ implies that the skewness of the probability
density function (PDF) of $p$, $S = \langle p^3\rangle /\langle
p^2\rangle ^{3/2}\,,$ is constant, suggesting that 
time irreversibility is robust and persists even in the limit
$\mathrm{Re}_\lambda \to \infty$.  It is important to stress that one might use
different dimensionless measures of the symmetry breaking, e.g., $
\tilde S = \langle p^3\rangle /\langle |p|^3\rangle\,,$ which directly
probes the ratio between the symmetric and asymmetric contributions to
the PDF. In the presence of anomalous scaling $S$ and $\tilde S$ can
have a different $\mathrm{Re}_\lambda$ dependence, as highlighted for the
problem of statistical recovery of isotropy \cite{BV2001}.

The aim of our work is twofold. First, we use direct numerical
simulations (DNSs) of 3D Navier-Stokes equations (NSEs) to quantify the
degree of recovery of time reversibility along single-particle
trajectories using different definitions as discussed above. Second,
we show that it is possible to extend the multifractal formalism (MF)
\cite{FP1985} to predict the scaling of the absolute value of the
Lagrangian power statistics.  Moreover, in order to explore a wider range of
Reynolds numbers, we also investigate the equivalent of the Lagrangian
power statistics in shell models \cite{biferale2003,bohr2005}.

The rest of the paper is organized as follows.
Section~\ref{sec:2} is devoted to a brief review of the multifractal formalism
for fully developed turbulence and the predictions for the statistics 
of the Lagrangian power. In Sec.~\ref{sec:3} we compare these 
predictions with the results obtained from direct numerical simulations
of the Navier-Stokes equations and from a shell model of turbulence.
Section~\ref{sec:4} is devoted to a summary and conclusions. The Appendix
reports some details of the numerical simulations.

\section{Theoretical predictions by the multifractal model}
\label{sec:2}
We start by recalling the MF for the
Eulerian statistics \cite{FP1985,Frisch1995}. The basic idea is to
replace the global scale invariance in the manner of Kolmogorov 
with a local scale invariance, by assuming that spatial velocity
increments $\delta_r u$ over a distance $r\ll L$ are characterized by
a range of scaling exponents $h\in \mathcal{I}\equiv(h_m,h_M)$,
i.e., $\delta_r u\sim u_L (r/L)^{h}$.  Eulerian structure functions
$\langle (\delta_r u)^q\rangle$ are obtained by integrating over
$h\in \mathcal{I}$ and the large-scale velocity $u_L$ statistics
$\mathcal{P}(u_L)$, which can be assumed to be independent of $h$. The MF
assumes the exponent $h$ to be realized on a fractal set of dimension
$D(h)$, so the probability to observe a particular value of $h$,
for $r\ll L$, is $\mathcal{P}_h(r) \sim (r/L)^{3-D(h)}$. Hence, we find
$\langle (\delta_r u)^q\rangle \sim \langle u_L^q\rangle \int_{h\in
  \mathcal{I}} dh (r/L)^{hq+3-D(h)} \sim \langle u_L^q\rangle
(r/L)^{\zeta_q}$, where a saddle-point approximation for $r\ll L$
gives
\begin{equation}
  \zeta_q= \inf_{h\in \mathcal{I}} \{hq+3-D(h)\}\,.
  \label{eq:zetaEMF}
\end{equation}
For the MF to be predictive, $D(h)$ should be
derived from the NSE, which is out of reach. One can, however,
use the measured exponents $\zeta_p$ and, by inverting
(\ref{eq:zetaEMF}), derive an empirical $D(h)$. 
Here, following \cite{SL1994}, we use
\begin{equation}
D(h)= 3-d_0-d(h)\left[\ln\left({d(h)}/{d_0}\right) -1\right]\,,
  \label{eq:dofh}
\end{equation}
with $d(h)={3(1/9-h)}/{\ln\beta}$ and $d_0={2}/{[3(1-\beta)]}$ corresponding,
via (\ref{eq:zetaEMF}), to
$\zeta_q=q/9+(2/3)(1-\beta^{q/3})/(1-\beta)$, which, for $\beta=0.6$,
fits measured exponents fairly well \cite{arneodo2008}.

The MF has been extended from Eulerian to Lagrangian velocity
increments \cite{borgas1993,boffetta2002}.  The idea is that temporal
velocity differences $\delta_\tau v$ over a time lag $\tau$, along
fluid particle trajectories, can be connected to equal time spatial
velocity differences $\delta_r u$ by assuming that the largest
contribution to $\delta_\tau v$ comes from eddies at a scale $r$ such
that $\tau \sim r/\delta_r u$.  This implies $\delta_\tau v \sim
\delta_r u$, with
\begin{equation}
\tau \sim T_L ({r}/{L})^{1-h} \, ,
\label{eq:ELbridge}
\end{equation}
where $T_L=L/u_L$.  By combining Eq.~(\ref{eq:ELbridge}) and the $D(h)$ obtained
from Eulerian statistics, one can derive a prediction for Lagrangian
structure functions, which has been found to agree with experimental
and DNS data
\cite{boffetta2002,chevillard2003,biferale2004,arneodo2008}.
The MF can be used also for describing the statistics of the 
acceleration $a$ along fluid elements \cite{borgas1993,biferale2004}.
The acceleration can be estimated by assuming 
\begin{equation}
a \sim {\delta_{\tau_\eta} v}/{\tau_\eta}\,.
\label{eq:acc}
\end{equation}
According to the MF, the dissipative
scale fluctuates as $\eta \sim (\nu L^h/u_L)^{1/(1+h)}$ 
\cite{FV1991}, which leads, via (\ref{eq:ELbridge}), to
\begin{equation}
  \tau_\eta \sim T (\nu/Lu_L)^{(1-h)/(1+h)}\,.
\label{eq:taueta}
\end{equation}
Substituting (\ref{eq:taueta}) in
(\ref{eq:acc}) yields the acceleration conditioned on given values of
$h$ and $u_L$:
\begin{equation}
  a \sim \nu^{(2h-1)/(1+h)} u_L^{3/(1+h)} L^{-3h/(1+h)}\,.
  \label{eq:acc2}
\end{equation}
Equation (\ref{eq:acc2}) has been successfully used to predict the
acceleration variance \cite{borgas1993} and PDF \cite{biferale2004}.

We now use
(\ref{eq:acc2}) to predict the scaling behavior of the Lagrangian
power moments with $\mathrm{Re}_\lambda$.  These can be estimated as $\langle
p^q \rangle\! \sim\!  \langle (au_L)^q\rangle\!
\!\sim\!\!  \int du_L
\mathcal{P}(u_L) \int_{h\in \mathcal{I}} dh \mathcal{P}_h(\tau_\eta)
(au_L)^q$ with
$\mathcal{P}_h(\tau_\eta)=(\tau_\eta/T)^{{[3-D(h)]}/{(1-h)}}$.  Using
(\ref{eq:taueta}) with $\nu=U_LL \mathrm{Re}_\lambda^2$ (with
$U_L^2=\langle u_L^2 \rangle$), we have
\begin{eqnarray}
\hspace{-0.5truecm}\frac{\langle p^q\rangle}{\epsilon^q}\! \sim\!\!  \int \!\!d\tilde{v}
\mathcal{P}(\tilde{v})\!\! \int_{h\in \mathcal{I}} \!\!\!\!\!\!\!dh  \tilde{v}^{[4q+h-3+D(h)]/(1+h)} \! \mathrm{Re}_\lambda^{2 [(1-2h)q-3+D(h)]/(1+h)}\,,
\label{eq:momp}
\end{eqnarray}
with $\tilde{v}=u_L/U_L$
\footnote{Possible divergences in $\tilde{v}\to 0$ should not be a concern as
  the MF cannot be trusted for small velocities}.  In the limit
$\mathrm{Re}_\lambda\to \infty$, a saddle point approximation of the integral
(\ref{eq:momp}) yields, up to a multiplicative constant (depending on
the large scale statistics), $\langle p^q\rangle/\epsilon^q\!\! \sim
\mathrm{Re}_\lambda^{\alpha(q)}$ with
\begin{equation}
  \alpha(q)=\sup_h\left\{2\frac{(1-2h)q-3+D(h)}{1+h}\right\}\,.
\label{eq:expo}
\end{equation}
\begin{figure}[t!]
\centering
\includegraphics[width=0.82\columnwidth]{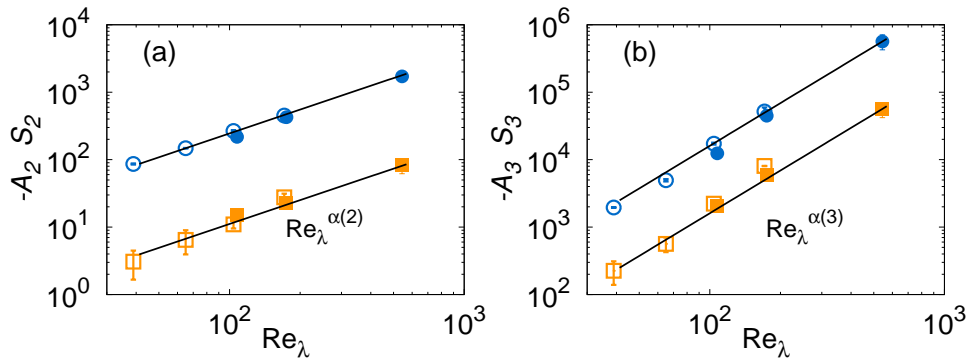}
\caption{Scaling behavior of Lagrangian power moments
  (\ref{eq:defmom}) $\mathcal{S}_q$ (blue circles) and
  $-\mathcal{A}_q$ (orange squares) for (a) $q=2$ and (b) $q=3$.  Data
  refer to DNS1 (closed symbols) and DNS2 (open symbols) datasets,
  described in the Appendix.  Solid lines show the
  slopes (a) $\alpha(2)=1.17$ and (b) $\alpha(3)=2.1$ predicted by the MF
  via (\ref{eq:expo}) with (\ref{eq:dofh}) for $\beta=0.6$.  Errors
  bars have been obtained as standard errors over independent
  configurations of the turbulent field. We used from 5 to 40
  configurations spaced by approximately $T_L$, depending on the resolution.
\label{fig:DNS1}}
\end{figure}

\section{Comparison with numerical simulations}
\label{sec:3}

To test the MF predictions (\ref{eq:expo}) we use two sets of DNS of
homogeneous isotropic turbulence on cubic lattices of sizes from
$128^3$ up to $2048^3$, with $\mathrm{Re}_\lambda$ up to $540$, obtained with
two different forcings (see the Appendix for
details). In particular, to probe both the symmetric and asymmetric
components of the Lagrangian power statistics, we study the nondimensional moments
\begin{equation}
\mathcal{S}_q=\langle
|p|^q\rangle/\epsilon^q, \quad \mathcal{A}_q=\langle
p|p|^{q-1}\rangle/\epsilon^q\,,
\label{eq:defmom}
\end{equation}
where the latter vanishes for a symmetric (time-reversible) PDF. In
Fig.~\ref{fig:DNS1} we show the second-and third-order moments of
(\ref{eq:defmom}) as a function of $\mathrm{Re}_{\lambda}$. 
We observe that (i) the MF prediction
(\ref{eq:expo}) is in excellent agreement with the scaling of
$\mathcal{S}_q$ (see also Fig.~\ref{fig:DNS2}) and (ii) the asymmetry
probing moments $\mathcal{A}_q$ are negative, confirming the existence
of the time-symmetry breaking, and scale with exponents compatible
with those of $\mathcal{S}_q$. This implies that time reversibility is
not recovered even for $\mathrm{Re}_\lambda \to \infty$.  Actually,
irreversibility is independent of $\mathrm{Re}_\lambda$ if measured in terms of
the homogeneous asymmetry ratio $\tilde S=\mathcal{A}_q/\mathcal{S}_q$, while if quantified in
terms of the standard skewness $S$, it grows as $\mathrm{Re}_\lambda^{\chi}$
with $\chi=\alpha(3)-(3/2) \alpha(2) \simeq 0.35$ due to anomalous scaling. 
In the
inset of Fig. {\ref{fig:DNS2} we compare $S$ with $\tilde S$.
Evaluating (\ref{eq:expo}) with $D(h)$ given by (\ref{eq:dofh}), we
obtain $\alpha(2)\approx 1.17$ and $\alpha(3)\approx 2.10$, which
are close to the $4/3$ and $2$ reported in \cite{xu2014}. We remark that
the authors of \cite{xu2014} explained the observed exponents by
assuming that the dominating events are those for which the particle
travels a distance $r \sim U_L\tau$ in a frozenlike turbulent
velocity field, so that $\delta_{\tau_\eta} v \sim (\epsilon
\tau_\eta U_L)^{{1}/{3}}$. Hence, for the acceleration
(\ref{eq:acc}) one has $a \sim
U_L^{{1}/{3}}\epsilon^{{1}/{3}}\tau_\eta^{-{2}/{3}}$, which, using the
dimensional prediction $\tau_\eta=(\nu/\epsilon)^{{1}/{2}}$, ends up
in $p\sim U_L a\sim
U_L^{{4}/{3}}\epsilon^{{2}/{3}}\nu^{-{1}/{3}}\sim \epsilon
\mathrm{Re}_\lambda^{2/3}$. This argument provides only a linear
approximation $2q/3$ for $\alpha(q)$, while the multifractal model
is able to describe its nonlinear dependence on $q$. In
Fig.~\ref{fig:DNS2} we show the whole set of exponents for both
$\mathcal{A}_q$ and $\mathcal{S}_q$ as observed in DNS data and 
compare them with the prediction (\ref{eq:expo}).
\begin{figure}[t!]
\centering
\includegraphics[width=0.7\columnwidth]{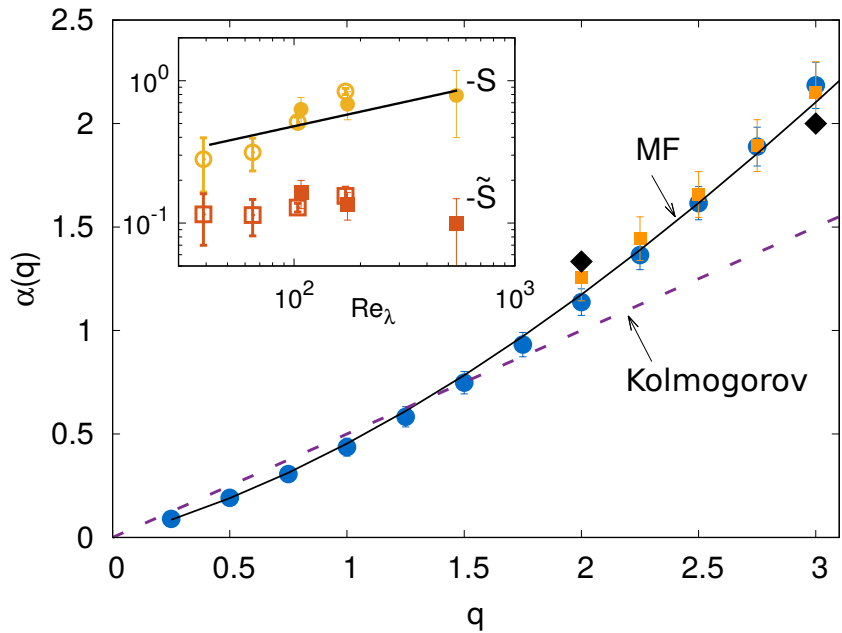}
\caption{Scaling exponents of Lagrangian power moments
  $\alpha(q)$ from DNS data, obtained by fitting $\mathcal{S}_q$
  (blue circles) and $-\mathcal{A}_q$ (orange squares) as power of
  $\mathrm{Re}_\lambda$. Error bars have been obtained by varying the fitting
  region; when they are not visible it is because they are of the order
  of or smaller than the symbol size.  Notice that $\mathcal{A}_q$ is
  positive for $q<1$, zero for $q=1$ (by stationarity) and negative
  for $q>1$.  We only show exponents for $q \geq 2$ because for $1 < q
  < 2$ insufficient statistics leads to a poor scaling behavior.
  Solid and dashed curves correspond to the MF (\ref{eq:expo}) and
  Kolmogorov [$\alpha(q)=q/2$] dimensional prediction, respectively. Black diamonds show the exponents found in
  \cite{xu2014}. The inset shows the nondimensional measure of the asymmetry in
  terms of the skewness $S=\langle p^3\rangle/\langle
  p^2\rangle^{3/2}$ (yellow circles) and of the statistically
  homogeneous asymmetry ratio $\tilde S=\langle p^3\rangle/\langle
  |p|^3\rangle$ (red squares). The solid line shows the slope
  $\alpha(3)-(3/2)\alpha(2) \simeq 0.35$ predicted by the MF (see the 
  text). open and closed symbols are as in Fig.~\ref{fig:DNS1}.
\label{fig:DNS2}}
\end{figure}

It is worth noticing that the MF provides an excellent prediction for the
statistics of $p$ also in 1D compressible turbulence, i.e., in the
Burgers equation, studied in \cite{grafke2015}. Here, out of a smooth
($h=1$) velocity field, the statistically dominant structures are
shocks ($h=0$). The velocity statistics is thus bifractal with
$D(1)=1$ and $D(0)=0$ \cite{bec2007}.  Adapting (\ref{eq:expo}) to one dimension
and noticing that $\mathrm{Re} \propto \mathrm{Re}_\lambda^2$, we have $\langle
p^q\rangle\sim \mathrm{Re}^{\alpha_{1D}(q)}$ with
$\alpha_{1D}(q)=\sup_{h}\{[(1-2h)q-1+D(h)]/(1+h)\}$, which for Burgers
means $\alpha_{1D}(q)=q-1$, in agreement with the results of
\cite{grafke2015}.

To further investigate the scaling behavior of
the symmetric and asymmetric components of the power statistics
in a wider range of  Reynolds numbers and with higher statistics,
in the following we study Lagrangian power  within the framework of shell
models of turbulence \cite{biferale2003,bohr2005}.
Shell models are dynamical systems built to reproduce the basic phenomenology of
the energy cascade on a discrete set of scales, $r_n=k_n^{-1}=L2^{-n}$
($n=0,\ldots,N$), which allow us to reach high Reynolds numbers.  For
each scale $r_n$, the velocity fluctuation is represented by a single
complex variable $u_n$, which evolves according to the differential equation \cite{Itamar}
\begin{equation}
\dot{u}_n=ik_n(u_{n+2}u^*_{n+1}-\frac{1}{4}u_{n+1}u^*_{n-1}
+\frac{1}{8}u_{n-1}u_{n-2})-\nu k_n^2 u_n+f_n\label{eq:sabra}
\end{equation}
whose structure is a cartoon of the 3D NSE in Fourier space but for the
nonlinear term that restricts the interactions to neighboring shells,
as justified by the idea localness of the energy cascade
\cite{rose1978}.  Energy is injected with rate
$\epsilon=\langle \sum_n \mathrm{Re}\{f_n u_n^*\}\rangle$.
See the Appendix for details on forcing
and simulations. As shown in \cite{Itamar}, this model displays anomalous scaling for
the velocity structure functions, $\langle |u_n|^q\rangle \sim
k_n^{-\zeta_q}$, with exponents remarkably close to those observed in
turbulence and in very good agreement with the MF prediction
(\ref{eq:zetaEMF}).

Following \cite{boffetta2002}, we model the Lagrangian velocity along
a fluid particle as the sum of the real part of velocity fluctuations
at all shells $v(t)\equiv \sum_{n=1}^{N} \mathrm{Re}\{u_n\}$.  Analogously, we
define the Lagrangian acceleration $a\equiv \sum_{n=1}^{N}
\mathrm{Re}\{\dot{u}_n\}$ and power $p(t)=v(t)a(t)$.
\begin{figure}[t!]
\centering
\includegraphics[width=0.94\columnwidth]{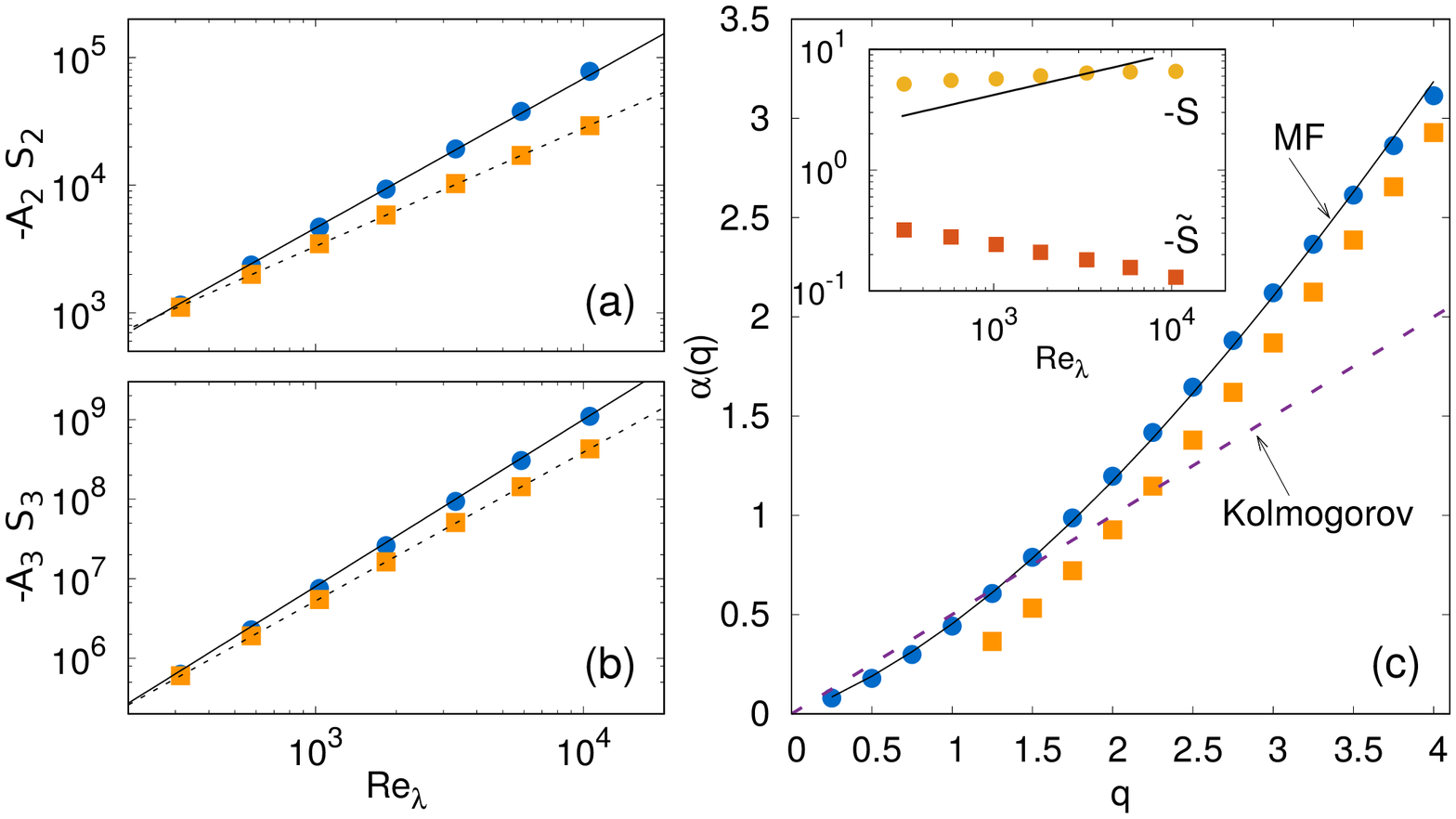}
\caption{Lagrangian power statistics in the shell model
  with $N=30$ shells at varying $\nu$.  $\mathrm{Re}_\lambda$-dependence of
  $\mathcal{S}_q$ and $-\mathcal{A}_q$ is shown for (a) $q=2$ and (b) $q=3$
  compared with the MF prediction (\ref{eq:expo}) (solid lines) and
  the best fit of the asymmetry-sensitive observables (dashed lines)
  providing slopes (a) $0.93(1)$ and (b) $1.87(1)$. Notice that
  $-\mathcal{A}_q$ is shifted upward to highlight the different
  scaling behavior.  (c) Scaling exponents $\alpha(q)$ obtained by
  fitting $\mathcal{S}_q$ (blue circles) and $-\mathcal{A}_q$ (orange
  squares) as power laws in $\mathrm{Re}_\lambda$, compared with (black solid
  curve) the MF prediction (\ref{eq:expo}) and (purple dashed curve)
  Kolmogorov dimensional scaling. Errors on the fitted values have been
  obtained by varying the fitting region; they are of the order of or
  smaller than the symbol size. The inset shows the nondimensional measure of the
  asymmetry in terms of the skewness $S=\langle p^3\rangle/\langle
  p^2\rangle^{3/2}$ (yellow circles) and of the statistically
  homogeneous asymmetry ratio $\tilde S=\langle p^3\rangle/\langle
  |p|^3\rangle$ (red squares).  Notice that the different scaling
  behavior of $\mathcal{S}_q$ and $-\mathcal{A}_q$ reflects on the
  $\mathrm{Re}_\lambda$ dependence of the $S$ that deviates from the MF slope
  $\alpha(3)-(3/2)\alpha(2)$ (solid line).  Data in (a), (b) and the
  inset in (c) have been obtained by averaging over ten
  realizations, each lasting $10^6\,T_L$; the standard error over the
  ten realization is of the order of or smaller than the symbol size.
  \label{fig:SM}}
\end{figure}
In Figs.~\ref{fig:SM}(a)  and \ref{fig:SM}(b) we show the moments $\mathcal{S}_q$ and
$\mathcal{A}_q$ for $q=2,3$ obtained from the shell model. The
symmetric ones $\mathcal{S}_q$ perfectly agree with the multifractal
prediction obtained using the same $D(h)$, i.e., (\ref{eq:dofh}) for
$\beta=0.6$, which fits the Eulerian statistics. The
asymmetry-sensitive moments $\mathcal{A}_q$ are negative (for $q>1$),
as in Navier-Stokes turbulence, and display a power-law dependence on
$\mathrm{Re}_\lambda$ with a different scaling respect to their symmetric analogs. In
particular, as summarized in Fig.~\ref{fig:SM}(c), we observe smaller
exponents with respect to the MF up to $q=4$. Rephrased in terms of the
skewness, these findings mean that the time asymmetry becomes weaker
and weaker with increasing Reynolds numbers if measured in terms of
$\tilde S$ [Fig. \ref{fig:SM}(c) inset], as distinct from what was observed for the NSE (Fig. \ref{fig:DNS2} inset). The standard
skewness $S$, on the other hand, is still an increasing function of
$\mathrm{Re}_\lambda$ though with an exponent smaller than the
MF prediction $\alpha(3)-(3/2)\alpha(2)$, because $\mathcal{A}_3$ has a
shallower slope  than the multifractal one.

\section{Conclusions}
\label{sec:4}
We have shown that the
multifractal formalism predicts the scaling behavior of the Lagrangian
power moments in excellent agreement with DNS data and with previous
results on the Burgers equation. In the range of explored
$\mathrm{Re}_\lambda$, we have found that symmetric and antisymmetric moments
share the same scaling exponents, and therefore the MF is able to
reproduce both statistics.  It is worth stressing that the effectiveness of the MF in
describing the scaling of $\mathcal{A}_q$ is not obvious  as the
MF, in principle, bears no information on statistical
asymmetries \footnote{See Sect.8.5.4 in \cite{Frisch1995} for a
  discussion.}. By analyzing the Lagrangian power
statistics in a shell model of turbulence, at Reynolds numbers much
higher than those achievable in DNS, we found that symmetric and
antisymmetric moments possess two different sets of exponents.  While
the former are still well described by the MF formalism, the latter,
in the range of $q$ explored, are smaller.  As a consequence, the
ratios $\mathcal{A}_q/\mathcal{S}_q$ in the shell model decrease with
$\mathrm{Re}_\lambda$. However, we observe that the mismatch between the two
sets of scaling is compatible with the assumption that $\mathcal{A}_q
\sim \mathcal{S}_q \langle \mathrm{sign}(p)\rangle$, i.e., that the main effect
is given by a cancellation exponent introduced by the scaling of $\mathrm{sgn}(p)$. Our findings raise the question whether the apparent
similar scaling among symmetric and asymmetric components in the NSE is
robust for large Reynolds numbers or a sort of recovery of
time symmetry would be observed also in Navier-Stokes turbulence as
for shell models.

We conclude by mentioning another interesting open question. In
\cite{xu2014,xu2014b} it was found that the Lagrangian power
statistics is asymmetric also in statistically stationary 2D
turbulence in the presence of an inverse cascade. Like in three dimensions,
the third moment is negative and its
magnitude grows with the separation between the timescale
of dissipation by friction (at large scale) and of energy
injection (at small scale), which is a measure of $\mathrm{Re}_\lambda$
for the inverse cascade range. Moreover, the scaling exponents are
quantitatively close to the 3D ones. This raises the question on the
origin of the scaling in 2two dimensions that cannot be rationalized within the MF,
since the inverse cascade is not intermittent
\cite{boffetta2012}. Likely, to answer the question one needs a better
understanding of the influence of the physics at and below the forcing
scale on the 2D Lagrangian power.

\begin{acknowledgments}
We thank F. Bonaccorso for computational support.  We acknowledge
support from the COST Action MP1305 ``Flowing Matter.''  L.B.
 and M.D.P. acknowledge funding from ERC under the EU $7^{th}$ Framework
 Programme, ERC Grant Agreement No. 339032. G.B. acknowledge
 Cineca within the INFN-Cineca agreement INF17\-fldturb.
\end{acknowledgments}

\appendix*
\section{Details on the numerical simulations}
\label{app1}

\subsection{Direct Numerical Simulations}
We performed two sets of DNSs at different 
resolutions and Reynolds numbers with two different forcing schemes.
The values of the parameters characterizing all the simulations are shown in Table \ref{table1}.
In all cases we integrated the Navier-Stokes equations
\begin{equation}
\partial_t {\bm u}  + {\bm u}\cdot {\bm \nabla} {\bm u} \equiv {\bm a} 
= - \bm {\bm \nabla} P + \nu \Delta {\bm u}  + {\bm f} \;, 
\label{eq:1}
\end{equation}
for the incompressible velocity field ${\bm u}({\bm x},t)$
with a fully parallel pseudo-spectral code, fully dealiased with $2/3$ rule \cite{orszag1971elimination}, in a cubic box 
of size $\mathcal{L}=2 \pi$ with periodic boundary conditions.
In (\ref{eq:1}) $P$ represents the pressure and
$\nu$ is the kinematic viscosity of the fluid.

For the set of runs DNS1 we used a Sawford-type stochastic
forcing, involving the solution of the stochastic differential
equations \cite{sawford1991}
\begin{equation}
\label{eq:sawford1}
\begin{cases}
d \tilde{f}_i = \tilde{a}_i(t) dt \, ,\\
d \tilde{a}_i = -a_1 \tilde{a}_i(t) dt -a_2 \tilde{f}_i(t) dt + a_3 dW_i(t)\, ,
\end{cases}
\end{equation}
where $a_1 = 1/\tau_f$, $a_2 = (1/8)/\tau_f^2$, $a_3 = \sqrt{2 a_1
  a_2}$, and $dW_i(t) = r \sqrt{dt}$ is an increment of a Wiener
process ($r$ is a random Gaussian number with $\langle r \rangle = 0$
and $\langle r^2 \rangle = 1$). The forcing $\bm{f}(\bm{k},t)$ in
Fourier space is then
\begin{equation}
\label{eq:sawford2}
\bm{f}(\bm{k},t) = 
\begin{cases}
i \bm{k} \times [i \bm{k} \times (0.16 \, k^{-4/3} \tilde{\bm{f}})] \quad &\text{for} \quad k \in [k_{f,\mathrm{min}}, k_{f,\mathrm{max}}] \, \\
0 \quad &\text{for} \quad k \notin [k_{f,\mathrm{min}}, k_{f,\mathrm{max}}] \, .
\end{cases}
\end{equation}
Time integration is performed by a second-order Adams-Basforth scheme with exact integration of the linear dissipative term \cite{canuto2006spectral}.

\begin{table}[ht!]
\begin{tabular}{|c|cccccccccccc|}
\hline
\hline
$\mathrm{Set}$ & $N$ & $\mathrm{Re}_{\lambda}$ & $\epsilon$ & $U$ & $L$ & $T_L$ & $\eta$ & $\tau_{\eta}$ & $T$ & $k_{f,\mathrm{min}}$ & $k_{f,\mathrm{max}}$ & $\tau_f$ \\
\hline

\hspace{0.05cm}DNS1\hspace{0.05cm} & $2048$ & $544$ & $1.43$ & $1.62$ & $4.51$ & $2.77$ & $0.0021$ & $0.015$ & $15$ & $0.5$ & $1$ & $0.14$\\
\hspace{0.05cm}DNS1\hspace{0.05cm} & $512$ & $176$ & $1.68$ & $1.74$ & $4.70$ & $2.70$ & $0.0083$ & $0.035$ & $10$ & $0.5$ & $1$ & $0.6$\\
\hspace{0.05cm}DNS1\hspace{0.05cm} & $256$ & $115$ & $1.19$ & $1.50$ & $4.26$ & $2.84$ & $0.019$ & $0.066$ & $48$  & $0.5$ & $1$ & $0.6$\\

\hline
\hspace{0.05cm}DNS2\hspace{0.05cm} & $1024$ & $171$ & $0.1$ & $0.529$ & $2.22$ & $4.19$ & $0.005$ & $0.063$ & $27$ & $0$ & $1.5$ & n/a\\
\hspace{0.05cm}DNS2\hspace{0.05cm} & $512$ & $104$ & $0.1$ & $0.520$ & $2.11$ & $4.06$ & $0.01$ & $0.10$ & $96$ & $0$ & $1.5$ & n/a\\
\hspace{0.05cm}DNS2\hspace{0.05cm} & $256$ & $65$ & $0.1$ & $0.513$ & $2.05$ & $3.98$ & $0.02$ & $0.16$ & $165$& $0$ & $1.5$ & n/a\\
\hspace{0.05cm}DNS2\hspace{0.05cm} & $128$ & $38.9$ & $0.1$ & $0.507$ & $1.95$ & $3.85$ & $0.04$ & $0.25$ & $165$& $0$ & $1.5$ & n/a\\
\hline
\hline
\end{tabular}
\caption{Type of forcing, resolution $N$, Reynolds number
  $\mathrm{Re}_{\lambda}=U \lambda/\nu$ [$\lambda=(5E/Z)^{1/2}$ is the Taylor
  microscale, $\epsilon$ the mean energy dissipation rate, $E$ the
  kinetic energy, and $Z$ the enstrophy], large-scale velocity $U=(2
  E/3)^{1/2}$, integral scale $L=U E/\varepsilon$, integral time
  $T_L=E/\varepsilon$, dissipative scale
  $\eta=(\nu^3/\varepsilon)^{1/4}$, Kolmogorov time
  $\tau_{\eta}=(\nu/\varepsilon)^{1/2}$, total time of integration
  $T$, and correlation time used in the forcing of DNS1 $\tau_f$ [see Eq.~(\ref{eq:sawford1})].
  Because of the different forcing in the two sets of
  simulations, for DNS2 the contribution of the modes at wave numbers
  $k \le 1$ have been removed in the analysis.
}
\label{table1}
\end{table}

For the set of runs DNS2 we use a deterministic forcing acting
on a spherical shell of wavenumbers in Fourier space 
$0 < |{\bf k}| \le k_f$, where 
$k_f=1.5$ with imposed energy input rate $\varepsilon$
\cite{lamorgese2005direct}. 
In Fourier space the forcing reads
\begin{equation}
\label{eq:2}
\bm{f}(\bm{k},t) = 
\begin{cases}
\varepsilon \bm{u}({\bf k},t) /[2 E_{f}(t)] \quad &\text{for} \quad k \in [k_{f,\mathrm{min}}, k_{f,\mathrm{max}}] \, \\
0 \quad &\text{for} \quad k \notin [k_{f,\mathrm{min}}, k_{f,\mathrm{max}}] \, .
\end{cases}
\end{equation}
where $E_{f}(t)=\sum_{k=0}^{k_f} E(k,t)$, and $E(k,t)$ is the energy
spectrum at time $t$. This forcing guarantees the constancy of the
energy injection rate. Notice that Eq.~(\ref{eq:2}) explicitly breaks
the time-reversal symmetry; however, owing to the universality
properties of turbulence with respect to the forcing, we expect this
effect to be negligible as compared to the energy cascade.  Time
integration is performed by a second-order Runge-Kutta midpoint method
with exact integration of the linear dissipative term
\cite{canuto2006spectral,boyd2001chebyshev}.  Simulations have a
resolution $N$ sufficient to resolve the dissipative scale with
$k_{\mathrm{max}} \eta \simeq 1.7$ ($k_{\mathrm{max}}=N/3$).  We have checked in the
simulations that the velocity field is statistically isotropic with a
probability density function (for each component) close to a Gaussian.

Simulations are performed for several large-scale eddy turnover times $T$,
after an initial transient to reach the turbulent state, in
order to generate independent velocity fields in stationary conditions. 
From the velocity fields 
the acceleration field is then computed by evaluating the right hand side of (\ref{eq:1})
and the power field is obtained as $p={\bm u} \cdot {\bm a}$.

\subsection{Simulations of the shell model}

As for the shell model (\ref{eq:sabra}), simulations have been
performed by fixing the number of shells $N=30$ and varying the
viscosity $\nu$ in the range $[3.16 \times 10^{-4} , 3.16 \times
  10^{-8}]$. For each value of $\nu$ we performed ten independent
realizations lasting approximately $10^6 T_L$ each.  Time integration is
performed using a fourth-order Runge-Kutta scheme with exact
integration of the linear term.  Forcing is stochastic and acts only
on the first shell $f_n=f\delta_{n,1}$.  The stochastic forcing is
obtained by choosing $f=F(f^R+if^I)$ with $F=1$ and
\begin{eqnarray}
  \dot{f}^{\alpha} &=& -\frac{1}{\tau_f}f^{\alpha} + \frac{\sqrt{2}}{\tau_f} \theta^\alpha(t) \label{eq:colored1} \, ,\\
  \dot{\theta}^{\alpha} &=& -\frac{1}{\tau_f}{\theta^{\alpha}} + \sqrt{\frac{2}{\tau_f}} \eta^\alpha(t) \, ,\label{eq:colored2}
\end{eqnarray}
where $\eta^\alpha$ is a zero mean Gaussian variable with correlation
$\langle \eta^{\alpha}(t)\eta^\beta(t')\rangle =
\delta_{\alpha\beta}\delta(t-t')$. As a result, $f^{\alpha}$ is a zero
mean Gaussian variable with correlation $\langle
f^{\alpha}(t)f^{\beta}(t')\rangle= \delta_{\alpha\beta}
\frac{1}{\tau_f}\exp(-|t-t'|/\tau_f) (|t-t'|+\tau_f)$.  In particular,
we used $\tau_f=1$, which is of the order of the large-eddy turnover
time $T_L$.  Using a constant amplitude forcing, we obtained, within
error bars, indistinguishable exponents (not shown).

%

\end{document}